\documentclass[twocolumn]{aastex631}             

\usepackage{graphicx}	
\usepackage{amsmath}	
\usepackage{amssymb}	
\usepackage{multirow}
\usepackage{soul}

\begin{document}

\title{Prospects for detecting cosmic filaments in Lyman-alpha emission across redshifts $z=2-5$}

\author[0009-0005-8855-0748]{Yizhou Liu}\thanks{E-mail: yzliu@nao.cas.cn}
\affiliation{National Astronomical Observatories, Chinese Academy of Sciences, Beijing 100101, China}
\affiliation{School of Astronomy and Space Science, University of Chinese Academy of Sciences, Beijing 100049, China}

\author{Liang Gao}\thanks{E-mail: lgao@bao.ac.cn}
\affiliation{School of Physics and Laboratory of Zhongyuan Light, Zhengzhou University, Zhengzhou 450001, China}
\affiliation{Institute for Frontiers in Astronomy and Astrophysics, Beijing Normal University, Beijing 102206, China}
\affiliation{National Astronomical Observatories, Chinese Academy of Sciences, Beijing 100101, China}

\author[0000-0001-7075-6098]{Shihong Liao}
\affiliation{National Astronomical Observatories, Chinese Academy of Sciences, Beijing 100101, China}

\author[0000-0002-2583-2669]{Kai Zhu}
\affiliation{Department of Astronomy, Tsinghua University, Beijing 100084, China}




%
\begin{abstract}

The standard $\rm \Lambda$CDM cosmological model predicts that a large amount of diffuse neutral hydrogen distributes in cosmic filaments, which could be mapped through Lyman-alpha (Ly$\alpha$) emission observations. We use the hydrodynamical simulation Illustris-TNG50 to investigate the evolution of surface brightness and detectability of neutral hydrogen in cosmic filaments across redshifts $z=2-5$. While the HI column density of cosmic filaments decreases with redshift, due to the rising temperature with cosmic time in filaments, the surface brightness of Ly$\alpha$ emission in filaments is brighter at lower redshifts, suggesting that the detection of cosmic filaments is more feasible at lower redshifts. However, most of the Ly$\alpha$ emission from cosmic filaments is around $10^{-21}$ $\rm erg\ s^{-1}cm^{-2}arsec^{-2}$, making it extremely challenging to detect with current observational instruments. We further generate mock images using the Multi-Unit Spectroscopic Explorer (MUSE) spectrograph installed on the Very Large Telescope (VLT) and a MUSE-like spectrograph on the upcoming Extremely Large Telescope (ELT). Our finding indicates that while the VLT can only detect filamentary structures made of dense gas in galactic centers, the ELT is expected to reveal much finer filamentary structures from diffuse neutral hydrogen outside of galaxies. Compared to the VLT, both the number density and the longest length of filaments are greatly boosted with the ELT. Hence the forthcoming ELT is highly promising to provide a clearer view of cosmic filaments in Ly$\alpha$ emission. 

\end{abstract}



\section{\bf Introduction} 
\label{sec:intro}

Cosmic filaments are important components of the cosmic web \citep{Bond1996}. Numerical simulations have demonstrated that cosmic filaments provide potential wells to trap gas, thus assisting star formation in their resident low mass dark matter halos \citep{Liao2019}. It has also been predicted by numerical simulations that a large amount of diffuse neutral hydrogen exists in cosmic filaments \citep{Popping2009}, particularly at high redshifts. The distribution of neutral hydrogen in filaments can be used to distinguish the nature of dark matter \citep{Gao2015}. Therefore it is of great importance to detect cosmic filaments traced by neutral hydrogen.  

Neutral hydrogen in filaments could be detected through Lyman-alpha (Ly$\alpha$) emission, however the detection is challenging due to their low brightness. If the UV background \citep{Haardt1996, Faucher2009, Haardt2012, Puchwein2019} is assumed to be the only ionizing radiation source, the Ly$\alpha$ surface brightness (SB) of self-shielded filaments is predicted to be approximate $\rm SB \approx 10^{-20} erg\ s^{-1}cm^{-2}arcsec^{-2}$ at redshift $z=3$ \citep{Gould1996}. This surface brightness is so faint for currently $\sim$8m class telescopes. Even after stacking 390 subcubes from very deep MUSE/VLT data, including the Hubble Deep Field South \citep[HDFS;][]{Bacon2010} and the MUSE Ultra Deep Field \citep[UDF;][]{Bacon2017}, with orientations determined by the position of neighbouring Ly$\alpha$ galaxies in the redshift range from $z=3$ to $z=4$, \citet{Gallego2018} found no detectable Ly$\alpha$ emission on intergalactic scales.

Detection becomes easier only when gaseous filaments are illuminated by more energetic sources like quasars or star-forming galaxies \citep{Cantalupo2005, Kollmeier2010}. The extended Ly$\alpha$ emissions surrounding quasars or galaxies indeed have been detected and catalogued, including Ly$\alpha$ blobs \citep{Matsuda2004, Nilsson2006, Francis2013, Martin2014, Ao2020}, Ly$\alpha$ nebulae \citep{Cantalupo2014, Hennawi2015, Borisova2016, Cai2017, Arrigoni_Battaia2018, Cai2019, Mukae2020, Daddi2021, Shimakawa2022, Erb2023, Li2023, Zhang2023} and Ly$\alpha$ halos \citep{Momose2014, Wisotzki2016, Leclercq2017, Erb2018, Kusakabe2022, Guo2024_a, Guo2024_b, Pharo2024}. These observed Ly$\alpha$ structures span from tens to hundreds of kiloparsecs. However, the detectability of extended and ultra low-luminosity Ly$\alpha$ emissions appearing as filamentary structures, called `Ly$\alpha$ filaments', remains limited. A recent study conducted by \citet{Bacon2021, Bacon2023} may represent the first detection of faint Ly$\alpha$ filaments at redshifts $z=3$ to $z=5$, called MUSE Extremely Deep Field (MXDF), including an extremely deep observation with an exposure time up to 150 hours (exceeding 140 hours in the central region but dropping to only about 10 hours in the outer regions). Therefore, the detection of Ly$\alpha$ filaments remains difficult and challenging.

A couple of theoretical studies have investigated the detectability of Ly$\alpha$ filaments at certain high redshifts \citep{Elias2020, Witstok2021, Byrohl2023, Renard2024, Tsai2024}. \citet{Elias2020} proposed that if the pixel resolution of MUSE/VLT images is degraded by a factor of $\simeq700$, diffuse gaseous filaments can be detected at $z=3$ at $5\sigma$ level. \citet{Byrohl2023} showed that filaments with extent $L\geq400$ $\rm pkpc$ and surface brightness $\rm SB\geq10^{-20}$ $\rm erg\ s^{-1}cm^{-2}arcsec^{-2}$ have a spatial number density of $\sim$$10^{-3}$ $\rm cMpc$ at $z=2$. \citet{Witstok2021} utilized a simulation that builds upon the Sherwood simulation project \citep{Bolton2017} to explore Ly$\alpha$ emission from the cosmic web at redshifts 3.6 and 4.8. They found that the observation of Ly$\alpha$ filaments within high-density regions at $z\leq4$ is promising. 

Most previous studies on this subject either focused on narrower redshift range or used relatively simpler galaxy formation models. In this paper, we employ a sophisticated galaxy formation simulation --- Illustris-TNG50 \citep{Nelson2019, Pillepich2019}, combined with our own Ly$\alpha$ emission calculation model, to investigate Ly$\alpha$ emission from the cosmic web across redshifts $2$ to $5$, which can be covered by current ground-based instruments, such as MUSE, Keck Cosmic Web Imager (KCWI) and Visible Integral-Field Replicable Unit Spectrograph (VIRUS). We focus on exploring the evolution of surface brightness and detectability of Ly$\alpha$ filaments in the above redshift range and discuss quantitatively the detectability of Ly$\alpha$ filaments with the current available $\sim$8m and forthcoming $\sim$30m class telescopes.

This paper is organized as follows: Section~\ref{sec:Methodology} introduces the TNG50 hydrodynamic simulation (\S~\ref{sec:TNG50}) and our theoretical method for calculating Ly$\alpha$ emission (\S~\ref{sec:modelling Lya emission}). Section~\ref{sec:Result} presents the redshift evolution of surface brightness (\S~\ref{sec:Redshift evolution of Lya filament}) and detectability (\S~\ref{sec:Observation}) of Ly$\alpha$ filaments across redshifts $z=2-5$. Section~\ref{sec:Discussion} and~\ref{sec:Conclusion} discuss and summarize our findings.


\section{\bf Methodology} 
\label{sec:Methodology}

\subsection{Cosmological simulation - TNG50} 
\label{sec:TNG50}

The TNG50 simulation \citep{Nelson2019, Pillepich2019} is a high-resolution magneto-hydrodynamical cosmological simulation of galaxy formation performed with the AREPO code \citep{Springel2010}. This simulation incorporates various key physical processes for galaxy formation, including photoionization heating from ultraviolet background (UVB) radiation, an additional local radiation field emitted by active galactic nuclei (AGN), and gas cooling through both primordial and metal-line mechanisms \citep{Weinberger2017, Pillepich2018}. The uniform and time-varying UVB is based on \citet{Faucher2009}, while the AGN ionization field is implemented following the prescription of \citet{Vogelsberger2013}. The AGN radiation feedback dominates over the UVB within 50 pkpc around galaxies \citep{Byrohl2021}, and its escape process is modeled following \citet{Hopkins2007}, where an obscuration factor is used to approximate the escape of AGN ionizing radiation under the assumption of optically thin gas. The ionization attenuation of dense gas that self-shields from UVB and AGN ionizing radiation is treated according to \citet{Rahmati2013}, with the collisional ionization and recombination modeled using \citet{Katz1996}. 


It is worth noting that the simulation includes only AGN and UVB as sources of ionization for the production of  Ly$\alpha$ photons and does not incorporate ionizing radiation from star formation. The star formation may provide a non-negligible contribution to the ionizing photon budget, which should be explored in future work.





In this study, we make use of the TNG50-3 simulation, which has an average gas mass resolution of $3.67\times10^{6}\ h^{-1}\mathrm{M_{\odot}}$, and a dark matter mass resolution of $1.96\times10^{7}\ h^{-1}\mathrm{M_{\odot}}$. The box size is 35 $h^{-1}\mathrm{cMpc}$. The simulation adopts the Planck cosmological parameters \citep{Planck2020}: $\Omega_{\rm m}=0.3089$, $\Omega_{\rm b}=0.0486$, $H_0=100h=67.74\ \mathrm{km\ s^{-1}Mpc^{-1}}$, $\sigma_8=0.8159$, and $n_{\rm s}=0.9667$.

\subsection{Modelling Ly$\alpha$ emission in simulation} 
\label{sec:modelling Lya emission}

\subsubsection{Ly$\alpha$ emission model} 
\label{sec:Lya emission model}

There are two distinct mechanisms that result in Ly$\alpha$ emissions. One avenue involves the recombination of an ionized hydrogen atom, produced by ultraviolet photons, with an electron, leading to the production of a Ly$\alpha$ photon. The Ly$\alpha$ luminosity density resulting from recombination can be expressed as:
\begin{equation}
    \epsilon_{\rm rec} = \alpha_{\rm B}(T)f_{\rm rec,B}(T)n_{\rm e}n_{\rm HII}E_{\rm Ly\alpha}.
	\label{eq:recombination emission rate}
\end{equation}
Here, $\alpha_{\rm B}(T)$ represents the recombination rate between an ionized hydrogen (HII) atom and an electron at temperature $T$ under the Case-B recombination, while $f_{\rm rec,B}(T)$ denotes the corresponding probability of producing a Ly$\alpha$ photon \citep{Draine2011, Dijkstra2014, Silva2016}. The variables $n_{\rm e}$ and $n_{\rm HII}$ correspond to the number densities of electrons and ionized hydrogen atoms, respectively. $E_{\rm Ly\alpha}$ refers to the energy of a Ly$\alpha$ photon.

The other avenue of Ly$\alpha$ photon production arises from the collision between a neutral hydrogen atom (HI) and an electron. The Ly$\alpha$ luminosity density arising from collisional excitation is described by the equation:
\begin{equation}
    \epsilon_{\rm coll} = q_{\rm coll}(T)n_{\rm e}n_{\rm HI}E_{\rm Ly\alpha},
	\label{eq:collision emission rate}
\end{equation}
where $q_{\rm coll}(T)$ represents the collisional excitation coefficient \citep{Scholz1990, Scholz1991, Dijkstra2014, Silva2016}, and $n_{\rm HI}$ denotes the number density of neutral hydrogen atoms. Note, in this paper, the total hydrogen number density, $n_{\rm H}$, is defined as the sum of neutral hydrogen, $n_{\rm HI}$, and ionized hydrogen, $n_{\rm HII}$.

\subsubsection{Analysis of radiation transfer effect \textnormal{\&} \\Conservative limitations} 
\label{sec:Conservative limitation}

The radiative transfer process of Ly$\alpha$ photons is very complex, and three processes need to be taken into account:

1) \textbf{Destruction by dust and $\rm H_{2}$:} The dust and $\rm H_{2}$ in a galaxy could destroy some of the Ly$\alpha$ photons produced by the galaxy, thereby reducing the Ly$\alpha$ emission from the galaxy and its surrounding medium. 

2) \textbf{Scattering process:} The typical astrophysical environments are optically thick (i.e., the optical depth $\tau > 1$) for Ly$\alpha$ photons \citep[$\tau \sim 10^{7}(N_{\rm HI}/10^{20}\rm cm^{-2})$, where $N_{\rm HI}$ is HI column density;][]{Dijkstra2019}. This implies that after being emitted, Ly$\alpha$ photons undergo multiple scattering events before eventually escaping from the surrounding optically thick medium. This process can smear out bright spots in the Ly$\alpha$ surface brightness map and cause Ly$\alpha$ filaments to appear fuzzy \citep{Byrohl2023}.

3) \textbf{Transmission rate along the line of sight (LoS):} For Ly$\alpha$ photons that escape from the optically thick region, the multiple scattering processes shift their frequencies to both sides of the line center. Thus, the spectrum of these escaped photons exhibits a symmetric double-peak structure, with higher HI column densities resulting in farther peak positions. The photons that escape toward the observer's LoS on the blue side may, due to Hubble expansion, gradually shift back to the line center and then be reabsorbed by HI along the LoS. This process could reduce the observed Ly$\alpha$ surface brightness.

While all these three processes influence the observed surface brightness of Ly$\alpha$ maps, the former two mainly impact the Ly$\alpha$ surface brightness of galaxies and their adjacent gas, which are not focus of this study. Therefore, we only consider the third process which will be described in detail in the following subsection. It should be noted that Ly$\alpha$ photons escaping from star-forming gas and AGNs are not included in our calculation. These photons may brighten galaxies and illuminate filaments in their vicinity \citep[e.g.,][]{Byrohl2023}, thereby increase their surface brightness on scales ranging from tens to hundreds of kiloparsecs. Additionally, resonant scattering redistributes Ly$\alpha$ photons across different spatial scales. This process affects the surface brightness of filaments in two opposing ways: (1) it can enhance the brightness by scattering photons from galaxies and circumgalactic medium (CGM); (2) it can reduce brightness by scattering photons into surrounding outer intergalactic medium (IGM). As demonstrated by \citet{Byrohl2023}, the overall effect of this complex process is an increase in the surface brightness of Ly$\alpha$ filaments. Since both the Ly$\alpha$ photons escaping from galaxies and the scattering process contribute to brighter Ly$\alpha$ structures, our estimation of surface brightness is conservative.


\subsubsection{Transmission rate} 
\label{sec:Transmission rate}

\begin{table} 
 \caption{Mean transmission rate at different redshifts.}
 \label{tab: the transmission rate at different redshift}
 \begin{tabular}{lllll}
  \cline{1-5}
  $z$ & 2 & 3 & 4 &  5\\
  \cline{1-5}
  Tr & 0.93 & 0.8 & 0.6 & 0.5\\
  \cline{1-5}
 \end{tabular}
\end{table}

We employ a Monte Carlo code \citep{Munirov2023} to calculate the spectrum of Ly$\alpha$ photons escaping from dense gas with $N_{\rm HI} \geq 10^{17} \mathrm{cm^{-2}}$, which dominates the Ly$\alpha$ emission from filaments. The transmission rate ($\rm Tr$) of escaping Ly$\alpha$ photons along the LoS is given by $\rm{Tr} = e^{-\tau}$, where the optical depth is defined as:
\begin{equation}
    \tau(\nu_{\rm em}) = \int_{0}^{\infty}n_{\rm HI}(s)\sigma_{\alpha}[\nu(s, \nu_{\rm em}),T] \mathrm{d}s.
	\label{eq:optical depth}
\end{equation}
Here, the distance traversed by the photon is denoted by $s$, with its frequency changing from $\nu_{\rm em}$ to $\nu=\frac{H(z)s}{c}\nu_{\rm em}$ during its propagation. The number density of neutral hydrogen atoms along the photon's path is assumed to be homogeneous and equal to the average value $n_{\rm HI}(z)$ at a given redshift $z$:
\begin{equation}
    n_{\rm HI}(s) = n_{\rm HI}(z) = x_{\rm HI}n_{\rm H}(z).
	\label{eq:number density of HI}
\end{equation}
Here, $x_{\rm HI}$ denotes the neutral fraction of hydrogen in the case of ionization equilibrium, $\Gamma_{\rm ion}n_{\rm HI}=\alpha_{\rm B}(T)n_{\rm e}n_{\rm HII}$, where $\Gamma_{\rm ion}$ indicates the ionization rate of the UV background. Under the optically thin approximation, the average value $n_{\rm HII}(z)$ is approximately equal to $n_{\rm H}(z)$, thus $x_{\rm HI} \approx \frac{\alpha_{\rm B}(T)n_{\rm e}}{\Gamma_{\rm ion}}$. Note, we also assume $n_{\rm e}=n_{\rm H}$, thus neglecting the influence of helium. This assumption is supported by \citet{Rahmati2013}, who demonstrated in Appendix D2 that it serves as a good approximation.


The cross-section for Ly$\alpha$ photons reads,
\begin{equation}
    \sigma_{\alpha}(x,T) = 5.9\times10^{-14} \mathrm{cm^{2}} \left(\frac{T}{10^{4}\rm K}\right)^{-1/2}\phi(x),
	\label{eq:cross-section of Lya}
\end{equation}
where $x=(\nu-\nu_{\alpha})/\Delta\nu_{\alpha}$ is a dimensionless frequency variable, with $\Delta\nu_{\alpha}=\nu_{\alpha}v_{\rm th}/c$, and $\phi(x)$ represents the Voigt function. We refer the readers to \citet{Dijkstra2019} for details.

We found that while Ly$\alpha$ photons escaping from dense gas with different HI column densities exhibit different symmetric double-peaked spectrum, their transmission rates remain approximately consistent. Assuming the IGM temperature is approximately $2\times10^{4}\rm K$ \citep{Schaye2001, Zaldarriaga2001}, we computed the specific transmission rate of escaping Ly$\alpha$ photons at each redshift, as shown in Tab. \ref{tab: the transmission rate at different redshift}, which is consistent with the mean transmission rate shown in Fig. 4 of \citet{Byrohl2020}. We include this transmission rate when calculating the Ly$\alpha$ surface brightness map. 


\subsubsection{Narrowband image generation} \label{sec:Narrowband image}

\begin{table} 
\newcommand{\tabincell}[2]{\begin{tabular}{@{}#1@{}}#2\end{tabular}}
 \caption{The observation parameters at different redshifts. $z$: redshift. R: angle resolution in $\rm arcsec$. P: pixel size in $\mathrm{kpc}$. S: sensitivity ($1\sigma, 1\rm arcsec^{2}$) in $10^{-20} \rm erg\ s^{-1}cm^{-2}arcsec^{-2}$. B: number of wavelength slices. ST: thick of the slice in $\mathrm{cMpc}$.}
 \label{tab: the parameters of observations}
 \begin{tabular}{llllll}
  \cline{1-6}
  $z$ & R & P & S & B & ST \\
  \cline{1-6}
  2 & 1 & 8.59 &     & 6 & 9.08 \\
  3 & 1 & 7.89 & 7.2 & 9 & 9.03 \\
  4 & 1 & 7.12 & 5.7 & 13 & 9.42 \\
  5 & 1 & 6.44 & 4.9 & 16 & 8.88 \\
  \cline{1-6}
 \end{tabular}
\end{table}

We compute the Ly$\alpha$ emission luminosity for non-star-forming gas in a given thin slice, then assign the luminosity of gas to a two-dimensional plane using the Cloud-In-Cell (CIC) method. A transmission rate is further applied to derive the final surface brightness map. Note, most of the surface brightness maps in this paper are generated with the parameters of the MXDF \citep{Bacon2021} unless otherwise noted. The parameters of this observation, including resolution (R), sensitivity (S) and number of wavelength slices (B) at different redshifts, are extracted from Lines 1, 16 and 21 of Tab. C.1 in \citet{Bacon2021}. We relist them in Tab. \ref{tab: the parameters of observations}. The observed narrowband width ($\Delta\lambda$) is $1.25 \times \rm B$\ \AA, where $1.25$\AA\ is the spectral line resolution (or wavelength slice) of MUSE. The slice thickness ($\Delta z$) of the narrowband image corresponding to the observed narrowband width is determined by 
\begin{equation}
    \Delta\lambda = \lambda_{\rm Ly\alpha}\Delta z, 
	\label{eq:cosmological broaden}
\end{equation}
where the rest wavelength of Ly$\alpha$ is $\lambda_{\rm Ly\alpha}=1215.67$\AA. The wavelength range of the MUSE spectrograph enables the observation of Ly$\alpha$ emission only in the redshift range from $z=2.8$ to $z=6.6$. Therefore, the number of wavelength slices at redshift $z=2$ in Tab. \ref{tab: the parameters of observations} was derived under the assumption of a slice thickness of approximately 9 $\rm cMpc$. The pixel size (P) and slice thickness (ST) are calculated from the angle resolution (R) and bandwidth (B), respectively. 

\begin{figure*}
\centering
\includegraphics[width=2\columnwidth]{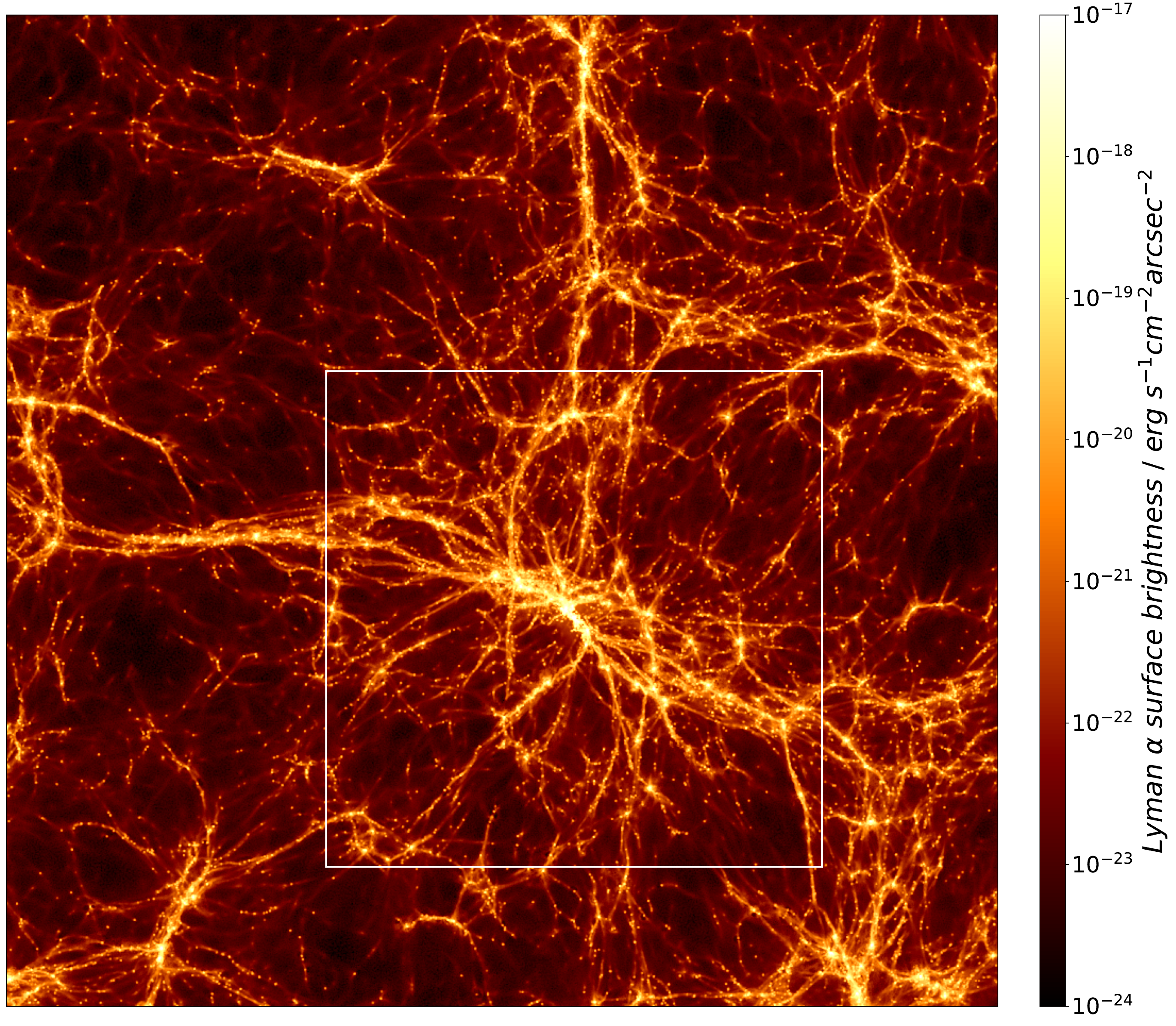}
\caption{Ly$\alpha$ surface brightness map for a slice from the TNG50 simulaition at $z=3$. This map highlights the existence of a large amount of neutral hydrogen in cosmic filaments, especially in the region marked by the white box, which will be explored in more detail in Fig.~\ref{fig:Lyman alpha redshift}, ~\ref{fig:density temperature and intensity statistic} and ~\ref{fig:Lyman alpha observation}.}
\label{fig:Lyman alpha sample}
\end{figure*}

\begin{figure*}
\centering
\includegraphics[width=2\columnwidth]{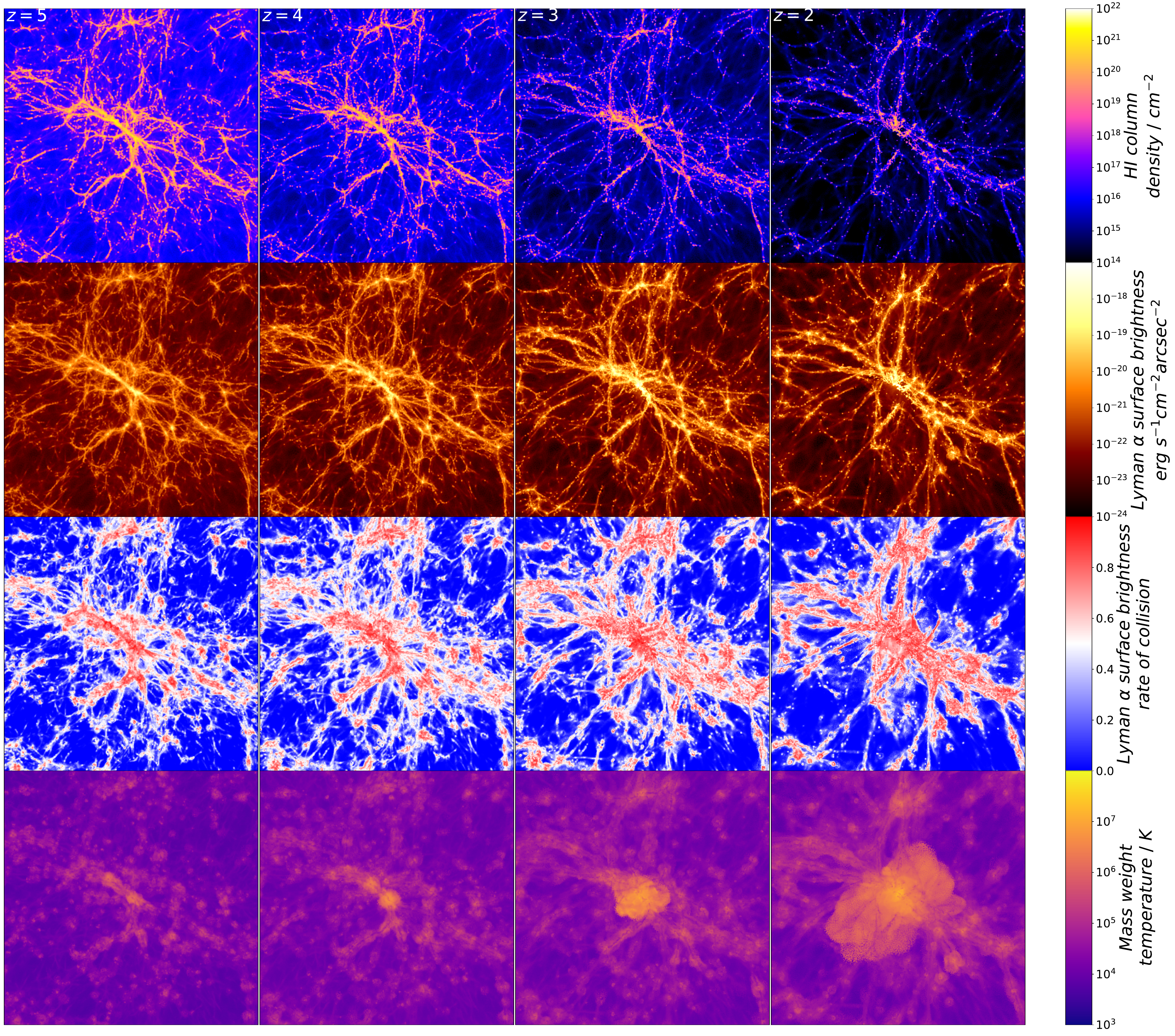}
\caption{The redshift evolution of HI column density, Ly$\alpha$ surface brightness, fraction of surface brightness due to collision excitation, and temperature map for the white enclosed region in Fig. \ref{fig:Lyman alpha sample} (from top to bottom); each panel has a scale of 10 $h^{-1}\rm cMpc$ on a side. From left to right, different columns show the maps at different redshifts.}
\label{fig:Lyman alpha redshift}
\end{figure*}

\begin{figure*}
\centering
\includegraphics[width=2\columnwidth]{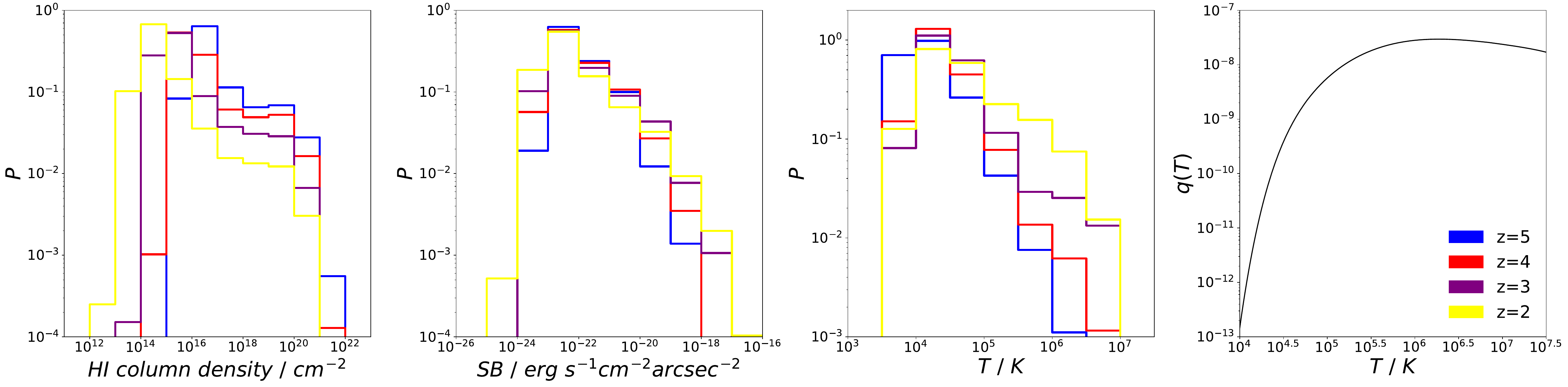}
\caption{The first three panels show the probability distribution of pixel values for HI column density, Ly$\alpha$ intensity and temperature maps in the Fig.~\ref{fig:Lyman alpha redshift}. The fourth panel shows the collisional excitation coefficient as a function of temperature $T$.}
\label{fig:density temperature and intensity statistic}
\end{figure*}

\begin{figure}
\centering
\includegraphics[width=\columnwidth]{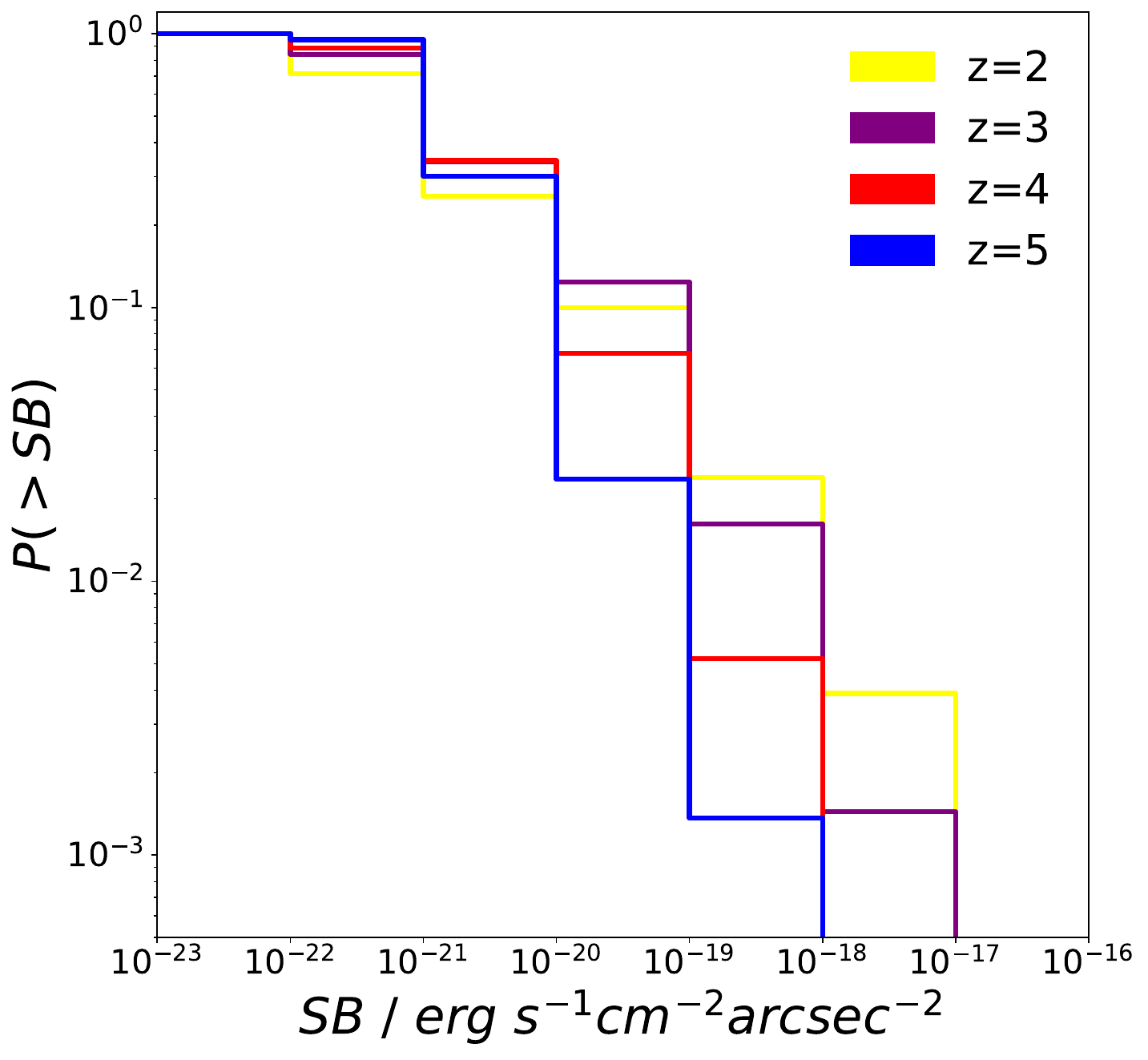}
\caption{The cumulative probability distribution function of pixel values in the Ly$\alpha$ surface brightness map of the entire simulation volume at different redshifts. Different redshifts are denoted by different colors, as indicated in the legend.}
\label{fig:Lyman alpha redshift statistics}
\end{figure}

\begin{figure*}
\centering
\includegraphics[width=2\columnwidth]{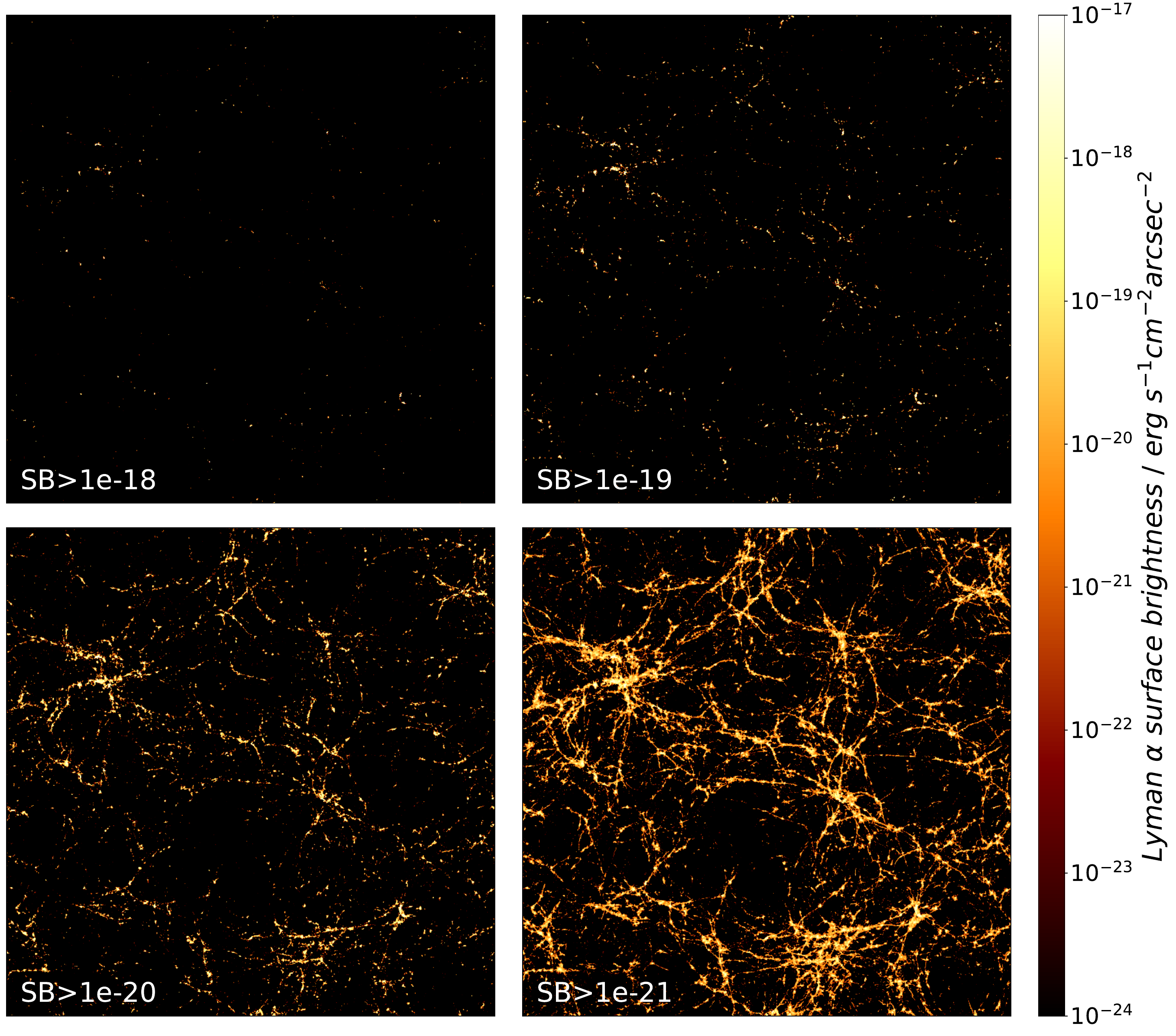}
\caption{The detectable Ly$\alpha$ structures in the entire simulation volume at different sensitivity levels at redshift $z=2$.}
\label{fig:Detectable Lyman alpha}
\end{figure*}

\begin{figure*}
\centering
\includegraphics[width=2\columnwidth]{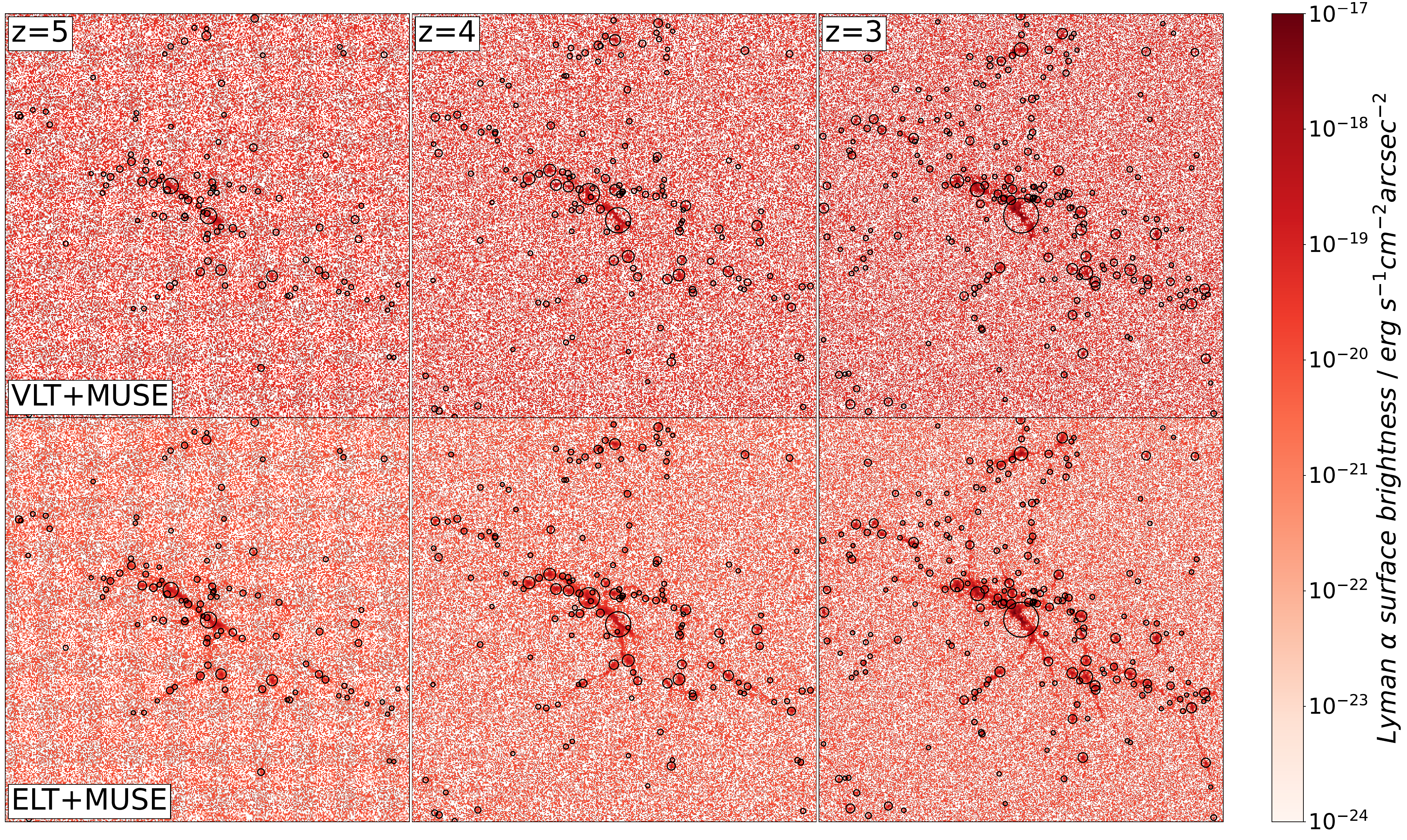}
\caption{Mock images of the Ly$\alpha$ surface brightness maps for the VLT (top) and ELT (bottom) at different redshifts. The black circles mark virial radius of all dark matter halos with mass $M > 10^{10}\ h^{-1}\mathrm{M_{\odot}}$.}
\label{fig:Lyman alpha observation}
\end{figure*}

\begin{figure}
\centering
\includegraphics[width=\columnwidth]{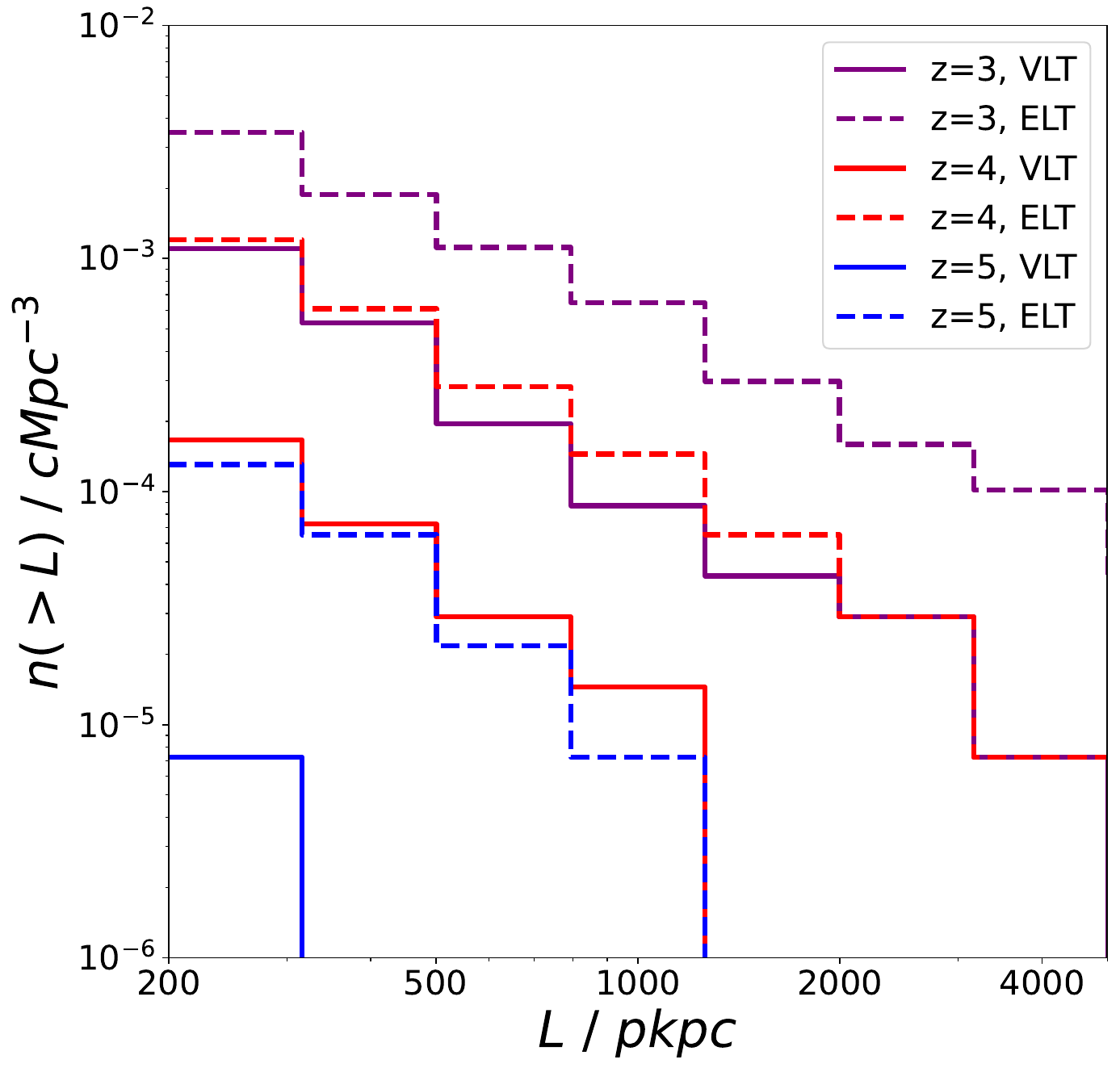}
\caption{The number densities of filaments with length greater than $L$ that can be detected by the VLT or ELT. The different color lines indicate the results at different redshifts, while the solid and dashed lines represent the results of the VLT and ELT, respectively.}
\label{fig:Filament statistics}
\end{figure}

\section{\bf Result} 
\label{sec:Result}

\subsection{Evolution of Ly$\alpha$ filaments in the simulation} 
\label{sec:Redshift evolution of Lya filament}

To obtain an overall visual impression of Ly$\alpha$ emission from the simulation, in Fig.~\ref{fig:Lyman alpha sample}, we present a Ly$\alpha$ intensity map for a slice extracted from the TNG50 simulation at $z=3$, with a thickness and angle resolution given by Tab. \ref{tab: the parameters of observations}. The map highlights the existence of a large amount of cosmic filaments illuminated by their resident neutral hydrogen. In order to see in more detail how the surface brightness of Ly$\alpha$ filaments evolves with redshift, we select a sub-region enclosed by the white square in the plot, which is rich of filamentary structures, and re-plot its HI column density, Ly$\alpha$ intensity map, surface brightness contributed by collision excitation as well as mass-weighted temperature across redshifts $z=2-5$ in Fig.~\ref{fig:Lyman alpha redshift}. 

As expected, the HI column density decreases monotonically with redshift. However, the redshift evolution of the Ly$\alpha$ surface brightness map does not follow the same trend. While the surface brightness of Ly$\alpha$ filaments generally exceeds $10^{-21}\rm erg\ s^{-1}cm^{-2}arcsec^{-2}$ across all redshifts and traces the HI filamentary structure at a given redshift, its surface brightness increases with decreasing redshift. This contradicts the simple expectation based on the relation between density and surface brightness,  
\begin{equation}
\begin{aligned}
I_{\rm rec} &\sim \frac{L_{\rm rec}}{(1+z)^{4}} \sim \frac{\epsilon_{\rm rec}}{(1+z)^{4}} \sim \frac{n_{\rm e}n_{\rm HII}}{(1+z)^{4}} \sim \frac{n_{\rm H}^{2}}{(1+z)^{4}} \\
&\sim (1+z)^{2}\ \rm and\\
I_{\rm col} &\sim \frac{L_{\rm col}}{(1+z)^{4}} \sim \frac{\epsilon_{\rm col}}{(1+z)^{4}} \sim \frac{n_{\rm e}n_{\rm HI}}{(1+z)^{4}} \sim \frac{n_{\rm H}^{3}}{(1+z)^{4}} \\
&\sim (1+z)^{5}.
\label{eq:redshift evolution of luminosity density}
\end{aligned}
\end{equation}
Here, under optically thin limit, the neutral fraction is approximately proportional to $n_{\rm H}$, thereby $n_{\rm HI} \sim n_{\rm H}^{2}$, while $n_{\rm e}$ and $n_{\rm HII} \sim n_{\rm H}$. The $n_{\rm H}$ evolve with redshift as $(1+z)^{3}$. As can be seen later, this contradiction is because the collisional excitation coefficient is highly sensitive to gas temperature. The third row of Fig.~\ref{fig:Lyman alpha redshift} presents the ratio of Ly$\alpha$ surface brightness from collisional excitation to the total Ly$\alpha$ surface brightness. Clearly, most of Ly$\alpha$ emission in filaments is dominated by collisional excitation, consistent with the findings of \citet{Witstok2021} and \citet{Byrohl2023}. The fourth row of the plot illustrates that the gas temperature of the filaments increases with redshift.

Fig.~\ref{fig:density temperature and intensity statistic} presents qualitative results of Fig.~\ref{fig:Lyman alpha redshift}. The first three panels show the probability distribution of pixel values for the HI column density, Ly$\alpha$ intensity and temperature maps in Fig.~\ref{fig:Lyman alpha redshift}, respectively. These qualitative results confirm our visual impression discussed above. The fourth panel shows the coefficient of collisional excitation as a function of gas temperature. Clearly, the collisional excitation is highly sensitive to temperature in the range $[10^{4}, 10^{6}]\rm K$. While the HI column density is highest at $z=5$, the rising gas temperature in filaments strongly enhances the collision-induced surface brightness of Ly$\alpha$ filaments, making them more detectable at lower redshifts. Note, a large temperature bubble presented at redshift 2 is caused by AGN feedback implemented in the simulation model.

In Fig.~\ref{fig:Lyman alpha redshift statistics}, we show the cumulative probability distribution function (hereafter CPDF) of pixel values obtained from the Ly$\alpha$ surface brightness map of the entire simulation volume. Results for different redshifts are distinguished by different colors. While the CPDFs show clear evolution with redshift, with brighter pixels increasing with redshift, more than $70$ percent of Ly$\alpha$ emission is below $10^{-21}$ $\rm erg\ s^{-1}cm^{-2}arsec^{-2}$ across the entire redshift range we consider here.

In order to illustrate the correspondence between different SB values and different structural sources, in Fig.~\ref{fig:Detectable Lyman alpha}, we show Ly$\alpha$ intensity maps from the entire simulation volume at a representative redshift $z=2$, above different thresholds (sensitivities). Apparently, the strongest emissions with SB $>10^{-18}$ $\rm erg\ s^{-1}cm^{-2}arsec^{-2}$ originate from relatively massive individual galaxies, which are more abundant at lower redshifts yet still very sparse in space. When the SB threshold decreases to $10^{-19}$ $\rm erg\ s^{-1}cm^{-2}arcsec^{-2}$, emissions from relatively low mass halos and surroundings of massive galaxies become detectable, therefore filamentary structures become visible. Further improvement to a sensitivity level of $10^{-20}$ $\rm erg\ s^{-1}cm^{-2}arcsec^{-2}$ allows the detection of emission from dense gas in filaments, revealing a clear filamentary structure in the Ly$\alpha$ intensity map. Again, detection is easier as redshift decreases for the two sensitivities. Finally, at a sensitivity level of $10^{-21}$ $\rm erg\ s^{-1}cm^{-2}arcsec^{-2}$, even some diffuse components in the filament become detectable, enabling the detection of striking filamentary features in the universe. In this case, detectability becomes more feasible at redshifts $3$ and $4$.

\subsection{Mock observation of Ly$\alpha$ filaments} 
\label{sec:Observation}

\subsubsection{Mock images} 
\label{sec:Mock image}

\begin{table} 
\newcommand{\tabincell}[2]{\begin{tabular}{@{}#1@{}}#2\end{tabular}}
 \caption{The sensitivity of MUSE installed on the VLT and ELT at different redshfits, assuming an exposure time of 150 hours. The unit of sensitivity (Sen.) is $10^{-20} \rm erg\ s^{-1}cm^{-2}arcsec^{-2}$.}
 \label{tab: the sensitivity of MUSE}
 \begin{tabular}{lllll}
  \cline{1-5}
  & $z$ & 3 & 4 & 5 \\
  \cline{1-5}
  VLT & Sen. & 7.2 & 5.7 & 4.9\\
  \cline{1-5}
  ELT & Sen. & 1.7 & 1.3 & 1.1\\
  \cline{1-5}
 \end{tabular}
\end{table}

To explore the detectability of cosmic filaments through real observations, we create mock images of Ly$\alpha$ emission for the MUSE spectrograph, which has been argued to be the most suitable instrument for observing the cosmic web across redshifts $z=3-5$ \citep{Witstok2021}. We consider both VLT and ELT telescopes. The VLT, which has a collecting area of 52 m$^2$, equipped with MUSE, has been utilized to detect a few diffuse Ly$\alpha$ emissions from filaments with an extremely deep 150 hours exposure time \citep{Bacon2021}. The ELT, with a collecting area of 978 m$^2$, is a forthcoming generation of optical telescope, and it is expected to significantly enhance the observation of Ly$\alpha$ filaments. We follow the methodology employed by \citet{Witstok2021} to calculate the sensitivity of MUSE installed on the ELT at different redshifts (assuming a $1/\sqrt{N}$ scaling of the noise level with $N$ times of collecting areas). The sensitivities of MUSE/VLT and MUSE/ELT are listed in Tab. \ref{tab: the sensitivity of MUSE}.

Fig.~\ref{fig:Lyman alpha observation} shows the mock images of the white box region in Fig.~\ref{fig:Lyman alpha sample} observed by MUSE/VLT and MUSE/ELT. Here we add Gaussian noise according to the sensitivity of both instruments. Each column corresponds to images generated at different redshifts. The black circles indicate sizes ($r_{200}$) of the dark matter halos with mass $M > 10^{10}\ h^{-1}\mathrm{M_{\odot}}$. 

As shown in the top row of the figure, only dense neutral hydrogen residing in the central regions of halos and subhalos can be seen with the VLT, while some of these halos distribute with a filamentary feature across all redshifts, albeit more structures can be seen at lower redshifts with the VLT. For the ELT (shown in the bottom row of the figure), benefiting from its high sensitivity, Ly$\alpha$ emissions from galaxies appear clearer, and the ELT can see emissions outside of the galaxies, particularly at redshifts $z\leq4$. These diffuse emission provide a clearer image of the filamentary structures of cosmic web. While even with ELT, at redshift 5, the detectable Ly$\alpha$ emission originating from diffuse neutral hydrogen remains elusive.

\subsubsection{Number density of detectable Ly$\alpha$ filaments} 
\label{sec:Number density}

To quantify the detectability of cosmic filaments with the VLT and ELT, we identify the filaments whose Ly$\alpha$ emission could be detected by these instruments, and further compute their number densities at various redshifts. The filament finder code used here was based on \citet[]{Liao2019} with some modifications. We first generate a Ly$\alpha$ surface brightness map for the entire simulation volume, then select pixels with surface brightness above the sensitivity of MUSE/VLT and MUSE/ELT (as listed in Tab. \ref{tab: the sensitivity of MUSE}), and finally use the Hoshen-Kopelman algorithm \citep{Hoshen1976} to link adjacent selected pixels as filaments. After identifying the filaments, we calculate their lengths using a method adapted from \citet{Cautun2014}. We compress all connected pixels of a filament to its central axis (i.e., to get the filament spine) and then measure the length of the filament spine (see Section 3.4 of \citet{Cautun2014} for details). In the following analysis, we only consider filaments with lengths $L > 200$ $\rm pkpc$.

In Fig.~\ref{fig:Filament statistics}, we present the number density of detectable filaments with different lengths. The solid and dashed lines correspond to results for the VLT and ELT, respectively. The lines with different colors distinguish the results for different redshifts. For the VLT, the detectable filament number densities at different redshifts approximate $7 \times 10^{-6}$ ($z=5$), $2 \times 10^{-4}$ ($z=4$), and $1 \times 10^{-3}$ ($z=3$) $\mathrm{cMpc}^{-3}$, with corresponding longest filament lengths of approximately $300$, $1000$, and $5000$ $\rm pkpc$, respectively. These results suggest that both the number densities and the maximal lengths of observable filaments increase as redshift decreases, consistent with the hierarchical assembly nature of the cold dark matter model. For the ELT, thanks to its high sensitivity, the number density of detectable filaments with length $L > 200$ $\rm pkpc$ increases greatly compared to the VLT, by factors of $20$, $8$, and $3$ times at redshifts $5$, $4$ and $3$, respectively. In addition, the maximal observed filament lengths also increase by large factors with the ELT. 

\section{\bf Discussion}
\label{sec:Discussion}

A few physical processes are not considered in this work, which may influence the surface brightness calculation. For instance, the escape of Ly$\alpha$ photons from galaxies (star-forming gas) and AGNs is not taken into account. These photons can brighten galaxies and illuminate filaments in their vicinity \citep[e.g.,][]{Byrohl2023}, thereby increasing their surface brightness on scales ranging from tens to hundreds of kiloparsecs. Consequently, the brightest regions (with surface brightness greater than $10^{-18}$ $\rm erg\ s^{-1}cm^{-2}arcsec^{-2}$) may appear even brighter, thus slightly expanding the areas of detectable regions in the Fig.~\ref{fig:Detectable Lyman alpha} and~\ref{fig:Lyman alpha observation}.

In addition, this study focuses on intrinsic emission from the IGM itself, as many previous studies \citep[e.g.,][]{Witstok2021} have done. However, \citet{Byrohl2023} demonstrated that the Ly$\alpha$ surface brightness of cosmic filaments could primarily be dominated by emission originating from galaxies and their surrounding gas halos. The scattering of these Ly$\alpha$ photons into the IGM  enhances the surface brightness of Ly$\alpha$ filament. As shown in the Figure 12 of \citet{Byrohl2023} that the enhancement of mean surface brightness in Ly$\alpha$ filament decreases as $\delta_{\rm DM}$ increases.  In the relevant regime for cosmic filaments ($\delta_{\rm DM}$ = [3 to 30]), above $\delta_{\rm DM} \sim 10$,  the enhancement from scattering of Ly$\alpha$ photons is quite small. Below the density, scattering dominate over the intrinsic emission by as large as a factor of several. Therefore, scattering primarily affects the outer parts of the Ly$\alpha$ filaments, while the central and denser regions remain largely unaffected. If these results are correct, they would certainly alter some of our results quantitatively, but not qualitatively. For instance, the filamentary structure would appear even more pronounced in the bottom right panel of the Fig.~\ref{fig:Detectable Lyman alpha} of our paper, however, the enhancement would be insufficient to be observational detectable even with the sensitivity of ELT/MUSE.

\section{\bf Conclusion}
\label{sec:Conclusion}

The standard cold dark matter model predicts a large amount of diffuse neutral hydrogen distributes in cosmic filaments, in particular at high redshifts. This diffuse neutral hydrogen can emit Ly$\alpha$ photons through recombination or collisional excitation processes, and therefore can potentially be mapped by some deep imaging observations, such as MXDF. In this paper, based on the hydrodynamical galaxy formation simulation TNG50, we develop a model to calculate Ly$\alpha$ emission from neutral hydrogen and investigate the evolution and detectability of Ly$\alpha$ emission from cosmic filaments spanning redshifts $2$ to $5$. Our results can be summarized as follows.

In the redshfit range considered in this work, the distribution of cosmic neutral hydrogen displays quite clear filamentary features. While the HI column density of the filamentary structures decreases with redshift, their Ly$\alpha$ surface brightness does not follow the same trend, in contrast to the simple expectation based on the relation between density and surface brightness. This is primarily driven by temperature-sensitive collisional excitation, which dominates the Ly$\alpha$ emission from filaments.

Over 70\% of Ly$\alpha$ emission from the simulation across all redshifts falls below $10^{-21}$ $\rm erg\ s^{-1}cm^{-2}arcsec^{-2}$, making it extremely difficult to detect with current observational facilities. However, as detection sensitivity improves from $10^{-18}$ to $10^{-20}$ $\rm erg\ s^{-1}cm^{-2}arcsec^{-2}$, filamentary structures become increasingly visible, particularly at lower redshifts. Sensitivity at the $10^{-21}$ $\rm erg\ s^{-1}cm^{-2}arcsec^{-2}$ level may reveal diffuse components of filaments, especially at redshifts $3$ and $4$. 

We generate mock images of Ly$\alpha$ emission based on the MUSE spectrograph installed on both the VLT and ELT to assess the detectability of cosmic filaments for current and forthcoming telescopes. The MUSE/VLT can only detect dense hydrogen in galactic centers, while the MUSE/ELT, with significantly higher sensitivity, is expected to see emission from IGM, particularly at redshifts $z\leq4$. However, detecting diffuse emissions at redshift $5$ remains challenging. 

We also calculate the number densities of detectable Ly$\alpha$ filaments with the VLT and ELT at different redshifts. Relative to the VLT, the ELT can detect $3$ to $20$ times more filaments and reveal longer filamentary structures across redshifts from 3 to 5. Particularly, the number density of detectable filaments reaches around $10^{-3} \ \rm cMpc^{-3}$ at redshift $z\leq4$. {The} upgrade detection capabilities of the ELT or other $\sim$30m class telescopes will likely provide a clear view of cosmic filaments in the future, thereby confirming a key prediction of $\rm \Lambda$CDM theory. Furthermore, since Ly$\alpha$ filaments are more easily detected at lower redshifts, BlueMUSE \citep{Richard2019} which is optimised for short wavelengths, is also well suitable for detecting Ly$\alpha$ filaments.

\section*{\bf Acknowledgments}
We are grateful to the anonymous referee for a very helpful referee report, which has significantly improved this work. We would like to thank Joris Witstok, Ewald Puchwein, Chris Byrohl and Bocheng Zhu for their help and discussions. We are also grateful to Simon D. M. White and Zheng Zheng for their useful discussions and comments. We acknowledge the supports from the National Natural Science Foundation of China (Grant No. 11988101) and the national Key Program for Science and Technology Research Development (2023YFB3002500). The IllustrisTNG simulations were undertaken with compute time awarded by the Gauss Centre for Supercomputing (GCS) under GCS Large-Scale Projects GCS-ILLU and GCS-DWAR on the GCS share of the supercomputer Hazel Hen at the High Performance Computing Center Stuttgart (HLRS), as well as on the machines of the Max Planck Computing and Data Facility (MPCDF) in Garching, Germany. KZ acknowledges the support from the Shuimu Tsinghua Scholar Program of Tsinghua University.

\software{matplotlib \citep{Hunter:2007},
          numpy \citep{harris2020array},  
          scipy \citep{Virtanen2020},
          h5py \citep{Collette2023}
          }

\bibliography{sample631}{}
\bibliographystyle{aasjournal}



\end{document}